\documentclass[12pt]{article}

\usepackage[psamsfonts]{amsfonts}

\newcommand{\no}{\noindent}

\newcommand{\be}{\begin{equation}} 
\newcommand{\bea}{\begin{eqnarray}}
\newcommand{\ee}{\end{equation}}
\newcommand{\eea}{\end{eqnarray}}
\newcommand{\non}{\nonumber}

\def\S{{\bf S}}
\def\R{{\bf R}}
\def\Z{{\bf Z}}
\def\C{{\bf C}}

\newcommand{\sm}{SU(3)\times SU(2)\times U(1)}

\begin{document}

\newpage
\renewcommand{\baselinestretch}{1.3}
\normalsize

\Large
\begin{center}
{\bf The String Landscape: On  Formulas for Counting Vacua}

\large

\vskip 1cm 
Tamar Friedmann$^*$ and Richard P. Stanley$^\dagger$
\vskip .6cm 
\normalsize
{\it $^*$University of Rochester, Rochester, NY}

{\it $^\dagger$Massachusetts Institute of Technology, Cambridge, MA}

\vskip .5cm 
\normalsize
\abstract{

We derive formulas for counting certain classes of vacua in the string/M theory landscape. We do so in the context of the moduli space of 
 M-theory compactifications on singular manifolds with $G_2$ holonomy. Particularly, we count the numbers of 
gauge theories with different gauge groups but equal numbers of $U(1)$ factors which are dual to each other. The vacua correspond to various symmetry breaking patterns of grand unified theories. 
Counting these dual vacua is equivalent to counting the number of conjugacy classes of elements of finite order inside Lie groups.  
We also point out certain cases where the conventional expectation is that symmetry breaking patterns by Wilson lines and Higgs fields are the same, but we show they are in fact different. 

}
\end{center}
\normalsize
\vskip .3cm


\section{Introduction}

\renewcommand{\baselinestretch}{1.3}
\normalsize

In the study of compactification of string or M theory on Calabi-Yau (CY) or $G_2$ manifolds, it was believed for many years that one could find a particular CY or $G_2$ manifold from which the theory resulting from  compactification would be the true vacuum: the Standard Model of particle physics. After a while, it became apparent that finding such a CY or $G_2$ manifold would be inordinately difficult due to the staggering number of possible manifolds, which form what is now known as the ''string landscape". The string landscape tied in with prior ideas related to the cosmological constant problem \cite{Bousso:2000xa, Feng:2000if, Brown:1987dd,Kachru:2003aw, Weinberg:1988cp}. There was then a paradigm  shift in the approach, from searching for a particular vacuum  towards counting the many vacua of the landscape \cite{Douglas:2003um}. 
An extensive discussion of the issue of counting vacua, following which there was 
 a surge of literature about statistics of vacua, appeared in \cite{Douglas:2003um} (see also \cite{Denef:2008wq} and references therein). There, in lieu of
a precise counting,  statistical methods  were used.

Some time before the paradigm shift of \cite{Douglas:2003um}, the first author took the counting approach and found the first precise formula for counting string or M-theory vacua   in the context of a study of the dynamics of M-theory compactified on singular spaces of $G_2$ holonomy that provide for gauge symmetry breaking  \cite{Friedmann:2002ct}.\footnote{For counting of supersymmetric vacua in gauge theories via the Witten index, see \cite{Witten:1982df, Witten:2000nv, Kac:1999gw,Selivanov:2002ii, de Boer:2001px} and references therein.}
The study of compactifications on singular manifolds was motivated by the realization that for smooth $G_2$ manifolds, compactification does not lead to any non-Abelian gauge theories \cite{Acharya:2004qe, Acharya:1998pm, Acharya:2000gb} and thus is not useful in obtaining the ultimate vacuum, i.e. the standard model of particle physics. 
However, when a manifold of  $G_2$ holonomy has an A, D, or E singularity - that is, a singularity of the form $\C ^2/\Gamma$ where $\Gamma$ is a finite, discrete subgroup of $SU(2)$ -- it is known via string dualities \cite{Hull:1995mz, Witten:1995ex,Atiyah:2001qf,Acharya:1998pm, Acharya:2000gb,Friedmann:2008rh} that the four-dimensional physics contains a supersymmetric non-Abelian gauge theory with gauge group given by the ADE Lie group corresponding to the ADE singularity. Therefore, the study of compactification of M-theory on singular $G_2$ spaces could lead to potentially realistic physical theories; for recent examples, see \cite{Friedmann:2003cd, Friedmann:2002ty,Acharya:2006ia, Acharya:2007rc, Bobkov:2009za}.

The space of supersymmetric M--theory vacua can be thought of as the quantum moduli space of M--theories compactified on $G_2$ manifolds. When the $G_2$ manifolds are  asymptotic to a cone over $(\S^3\times \S^3)/\Gamma$, it has been shown \cite{Atiyah:2001qf,Acharya:1998pm, Acharya:2000gb, Atiyah:2000zz} that the quantum moduli space is a Riemann surface that smoothly interpolates between three different classical limits: 
one of these classical limits corresponds to non-abelian supersymmetric ADE gauge theories that arise from ADE singularities in the $G_2$ manifold, while the two others correspond to theories on a smooth $G_2$ manifold which does not admit any normalizable zero modes. Since   theories with no normalizable zero modes have a mass gap,  the smooth interpolation between such theories and non-abelian supersymmetric gauge theories has been viewed as evidence supporting the statement that non-abelian supersymmetric gauge theories have a mass gap  \cite{Acharya:2000gb, Atiyah:2001qf, Atiyah:2000zz, Acharya:2004qe}. 

This picture may be generalized to $G_2$ manifolds that are asymptotic to  a cone over  $(\S^3\times \S^3)/\Gamma$ where $\Gamma$ is a larger  group such as  $\Gamma = \Gamma _1 \times \Gamma _2 \times \Gamma_3 \in [SU(2)]^{ 3}$, where each $\Gamma _i$ is a finite discrete subgroup of $SU(2)$.  In this case,  it was shown in \cite{Friedmann:2002ct} that the three classical limits  appearing in the moduli space correspond to supersymmetric ADE gauge theories that admit a natural gauge symmetry breaking by Wilson lines\footnote{The idea of symmetry breaking by Wilson lines dates back to the 1980's \cite{hosotani, goodwit, CHSW, witSB} and  has appeared and re-appeared in the string/M/F theory literature. For some recent examples see \cite{Friedmann:2002ct,Friedmann:2003cd,Gukov:2003cy,arXiv:0802.2969,arXiv:0802.3391,arXiv:1107.0733}.} of a gauge group $G$  of ADE type to a subgroup $H$.
The moduli space then consists of disconnected branches  classified by the number of $U(1)$ factors that appear in the subgroup $H$ of $G$ that remains after the gauge symmetry breaking. As in \cite{Friedmann:2002ct}, we denote the branch with $s-1$ $U(1)$ factors by ${\cal N} _{s, \Gamma}$. The number of $U(1)$ factors also corresponds to the number of zero modes that appear in the theory. 
One example stands out: when $\Gamma = \Z_5\times \Z _q$ and $q$ is prime to 5, we obtain the first manifestation via M--theory of Georgi-Glashow grand unification: an $SU(5)$ grand unified gauge group is broken by Wilson lines precisely to the gauge group of the standard model $\sm$ \cite{Georgi:1974sy,Friedmann:2002ct,Friedmann:2003cd}. Having a single $U(1)$ factor, this theory appears on the branch ${\cal N} _{2, \Z_5\times \Z _q}$. 

The smooth interpolation between any two points on a given branch can be interpreted as a duality betwen theories with different gauge groups $H$ and $H'$ but equal number of $U(1)$ factors. For example, if the $G_2$ space has an $A_{n-1}$ singularity, we have an $SU(n)$ gauge theory which can be broken to $s$ non-abelian subgroups and $s-1$ abelian $U(1)$ factors via
\[ SU(n)\longrightarrow \Pi _{i=1}^s SU(n_i)\times U(1)^{s-1}~,
\]
where $\sum n_i = n$. For fixed $s$, all these theories appear on a single branch and are interpreted as dual to each other. This same duality, originally obtained in \cite{Friedmann:2002ct} via M-theory compactifications, was later rediscovered in \cite{Cachazo:2002zk} via field theoretic methods.

The question now becomes, how many such theories, or -- equivalently -- semi-classical vacua, actually appear on each branch ${\cal N} _{s,\Gamma}$ of the quantum moduli space? Unlike the more general case of \cite{Douglas:2003um}, where counting is not possible, a precise formula in the particular case of $A_n$ singularities actually was presented already in \cite{Friedmann:2002ct} (without the derivation).
 As it happens, this formula constituted the first actual counting of vacua with gauge symmetry breaking in the string/M theory landscape context, and the first author has received requests for the derivation.

 It is the purpose of this paper to provide the counting formulas not only for $A_{n-1}$ singularities corresponding to the grand unified group $G=SU(n)$, but also to provide analogous formulas for $G=SO(n)$, $Sp(n)$, $U(n)$, and $O(n)$. The prudent reader would notice that these are the A, B, C, and D Lie groups, while the M-theory compactification we started with  corresponds to gauge theories with A, D, or E gauge groups only. While the formulas we derive  for A and D groups apply directly to the given M-theory compactifications on A and D singularities, the formulas for B and C groups do not seem at first sight to be directly related to any such compactification. However,  we note that one may obtain the Dynkin diagrams corresponding to the B and C groups by folding the A or D Dynkin diagrams. Specifically,  $B_n$ may be obtained by folding $D_{n+1}$, and $C_n$ may be obtained by folding $A_{2n-1}$ \cite{stekol}.  Therefore,  our formulas for B and C groups do apply  to compactifications that allow for such folding action. 
Also, the formulas we obtain tell us new information about symmetry breaking patterns for grand unified theories (GUTs) that may be useful beyond the M-theory compactifications from which we obtained them. 

In all the cases under consideration, the locus of the singularity in the manifold of $G_2$ holonomy is a cycle which has a non-trivial fundamental group, and it is the fundamental group which allows for the GUT symmetry breaking by Wilson lines. The same Wilson lines naturally induce a fractional three-form flux that can stabilize moduli
\cite{Gukov:2003cy}, an important consideration when attempting to obtain standard model physics from string or M theory.

Note that the vacua we consider are supersymmetric; to reach the landscape of non-supersymmetric vacua, one needs to put in a supersymmetry breaking mechanism. Also, note that the specific manifolds of $G_2$ holonomy we consider are not compact; our discussion is relevant to M-theory on compact manifolds of $G_2$ holonomy with ADE singularities whose locus is a 3--cycle with non-trivial fundamental group \cite{Atiyah:2001qf, Friedmann:2002ct}.

This paper is organized as follows. In Section \ref{vacconj} we explain the connection between vacua and conjugacy classes. In Sections \ref{unitary}, \ref{symplectic}, and \ref{orthogonal} we derive the counting formulas  for the unitary and special unitary groups, symplectic groups, and   orthogonal  groups, respectively.
In Section \ref{wilsonvshiggs} we point out a subtle difference in symmetry breaking patterns by Wilson lines and by adjoint Higgs fields.  In a separate manuscript \cite{mathy}, 
the counting is addressed from a purely mathematical approach. 
 We leave for future work the analogous counting in the exceptional compact Lie groups as well as in non-compact Lie groups. 

Two of the quantities we derive, Eqns. (\ref{qinsupder}) and (\ref{qinspnder}), were obtained in \cite{81h:20052} using the full machinery of Lie structure theory. Our methods are combinatorial and apply not only to simply connected or adjoint groups as do the methods of  \cite{81h:20052}, so we can obtain results about $O(n)$, $SO(n)$, and $U(n)$ in addition to $SU(n)$ and $Sp(n)$.

\section{Vacua as conjugacy classes}\label{vacconj}


In this section we show that  counting vacua in the string/M theory landscape can be reformulated in terms of  counting certain  conjugacy classes in Lie groups. 

 Consider manifolds of $G_2$ holonomy that are asymptotic to a cone over $(\S ^3 \times \S ^3)/(\Z_p\times \Z_q)$, with $(p,q)=1$. The manifold $\S^3 \times \S^3$ can be described \cite{Atiyah:2001qf, Friedmann:2002ct} as a homogeneous space $SU(2)^3/SU(2)$ where the equivalence relation is 
\[ (g_1, g_2, g_3)\sim (g_1h, g_2h, g_3h), \; \; \; g_i, h\in SU(2).
\]
A cone over this manifold may be obtained by ''filling in" one of the three $SU(2)$ factors, that is, by allowing say $g_1\in {\bf B}^4$ (a ball of radius 1 in four dimensions) rather than just $g_1\in \S^3 \sim SU(2)$. 
The action of $\gamma \times \delta \in \Z_p\times \Z_q$ is then given by 
\[ (g_1, g_2, g_3)\mapsto (\gamma g_1, \delta g_2, g_3).\]
As shown in \cite{Friedmann:2002ct}, when $(p,q)=1$ the compactification of M-theory on such manifolds leads to semi-classical vacua that are either an $SU(p)$  gauge theory on a seven-dimensional locus of the form $\R ^4\times (\S^3/\Z_q)$, or  an $SU(q)$ gauge theory on $\R ^4\times (\S^3/\Z_p)$.\footnote{The requirement that $q$ and $p$  are relatively prime ensures that the seven-dimensional loci are not singular.} These gauge theories are the vacua that we will be counting in this paper. 

We may also consider manifolds of $G_2$ holonomy that are asymptotic to a cone over $(\S ^3 \times \S ^3)/(\Gamma _1 \times \Gamma _2)$, where now the $\Gamma _i$ are finite subgroups of $SU(2)$ that are not necessarily cyclic, i.e. they may be of D or E type as well. We still require $(|\Gamma _1|, |\Gamma _2|)=1$ where  $|\Gamma _i|$ denotes the number of elements in $\Gamma _i$.\footnote{Again, this condition ensures the smoothness of the seven-dimensional locus of the singularity.} Since the DE groups all have even order, at least one of the $\Gamma _i$ must be of A type, so we consider $(\S ^3 \times \S ^3)/(\Gamma \times \Z_q)$ with $\Gamma$ an ADE group such that $(|\Gamma |, q)=1$. We obtain vacua that are certain combinations of $SU(p)$, $SO(2n)$, and $E_{6,7,8}$ gauge groups living on cycles with (binary) cyclic, dihedral, tetrahedral, octahedral, and icosahedral fundamental groups. 

All the vacua we will be counting are gauge theories living on a manifold $M$ with non-trivial fundamental group $\pi _1(M)$ (in the above example, $M$ was $\R ^4\times (\S^3/\Z_p)$ or $\R ^4\times (\S^3/\Z_q)$ and $\pi _1(M)$ was $\Z_q$ or $\Z_p$). Therefore, these vacua admit Wilson lines (flat connections), that is, homomorphisms  of the fundamental group into the gauge group, $U:\pi_1(M)\rightarrow G$, given by \cite{gsw}:
\[U_\gamma = P \exp \int_\gamma A dx \]
where $\gamma$ is a non-trivial loop in $M$ and $A$ is a flat connection. 
In the presence of a Wilson line, the gauge group $G$ is broken to the subgroup  $H_\gamma \subset G$ that commutes with $U_\gamma$: 
\[ H_\gamma =\{ h\in G \; \; | \; \; h U_\gamma h ^{-1}=U_\gamma  \} ~.\]
For any other element $\gamma '$ of the fundamental group for which $U_{\gamma '}$  is conjugate in the gauge group to $U_\gamma$, the remaining subgroup $H_{\gamma '}$ will be conjugate to $H_\gamma $ and isomorphic to it.  We will be counting the theories that have distinct symmetry breakings, so we count each conjugacy class of Wilson lines only once. For example, when the fundamental group is  $\Z_q$ and the gauge group is $SU(p)$, a conjugacy class of Wilson lines is simply a  conjugacy class consisting of elements of $SU(p)$  order dividing $q$, i.e. elements $x\in SU(p)$ such that $x^q=I$.

\section{Counting vacua associated with unitary  groups}\label{unitary}
\renewcommand{\baselinestretch}{1.3}
\normalsize

In this section, we  derive the formulas for the number of semiclassical vacua that appear in the moduli space of M-theory compactifications on singular spaces of $G_2$ holonomy that are asymptotic to a cone over $(\S ^3 \times \S ^3)/\Z_p\times \Z_q$. 

Based on the previous section, this number is equal to the number of conjugacy classes of elements of order dividing $q$ in $SU(p)$ (i.e. $x^q=1$, $x\in SU(p)$) where $p$ and $q$ are relatively prime (or the same with $p$ and $q$ exchanged). In the process, we also count the number of such conjugacy classes in the full unitary group $U(p)$, where there are no constraints on $p$ and $q$. 

Since every conjugacy class of $U(p)$ or $SU(p)$ contains diagonal matrices, the problem reduces to counting inequivalent diagonal matrices. For both $U(p)$ and $SU(p)$, the diagonal entries are $q$th roots of unity, $e^{2\pi i k/q}$ with $k=0, 1, \ldots , q-1$; for $SU(p)$, we have the additional condition that their product - the determinant of the matrix - must be unity. 

We proceed through an example. Let $p=7$ and $q=4$. Partition $p$ into $q$ parts corresponding to the $q$ possible values for the diagonal entries, keeping in mind that  not all $q$ roots of unity are required to appear. Consider the following diagram:

\begin{center}
\begin{picture}(100,1)  
\put(0,1) {\circle*{4.0}} 
\put(10,1) {\circle*{4.0}} 
\put(20,1) {\circle*{4.0}} 
\put(30,1) {\circle*{4.0}} 
\put(40,1) {\circle*{4.0}} 
\put(50,1) {\circle*{4.0}} 
\put(60,1) {\circle*{4.0}} 
\put(70,1) {\circle*{4.0}} 
\put(80,1) {\circle*{4.0}} 
\put(90,1) {\circle*{4.0}} 
 \put(20,1){\circle{11.0}}
 \put(60,1){\circle{11.0}}
 \put(70,1){\circle{11.0}}
\end{picture}
\end{center}
This diagram corresponds to the partition $\{n_k\}=( n_0, n_1, n_2, n_3)=(2,3,0,2)$, $\sum n_k = p$. This partition corresponds to the matrix diag$(\exp{\frac{2\pi i}{4}}(0,0,1,1,1,3,3)) =$ diag$(1,1, e^{\frac{2\pi i}{4}}, e^{\frac{2\pi i}{4}}, e^{\frac{2\pi i}{4}}, e^{\frac{6\pi i}{4}}, e^{\frac{6\pi i}{4}})$, which breaks $U(7)$ to $U(2)\times U(3)\times U(2) = SU(2)\times SU(3)\times SU(2) \times U(1)^3$. There is a bijection between such diagrams and inequivalent diagonal matrices in $U(7)$. Note that the exponents of the entries in the corresponding matrix are nondecreasing, so there is no danger of over-counting any equivalent diagonal matrices. 

The above diagram has $p+q-1$ dots, with $q-1$ of them circled. The number of distinct such diagrams is given by 
\be \label{qinupder} {p+q-1 \choose q-1} = {p+q-1 \choose p},\ee
and this is the number of conjugacy classes of elements of order $4$ or $2$ in $U(7)$, or more generally the number of conjugacy classes of elements of order $q$ or dividing $q$ in $U(p)$. This number is also the number of "weak $q$-compositions of $p$" \cite{stanley}.

Now impose $(p,q)=1$ and turn to $SU(p)=SU(7)$. We must take into account the requirement that the determinant of the matrix must be unity. Consider the following set of $q=4$ distinct partitions related to each other by cyclic permutations, their corresponding $U(7)$ matrices, and the sums of the diagonal exponents: 
\bea \non
\{n_k^{(0)}\}&=&(2,3,0,2)\hskip 1cm \mbox{diag}(\exp{\frac{2\pi i}{4}}(0,0,1,1,1,3,3)) \hskip 1cm  \sum kn_k^{(0)} \equiv 1 \mbox{ mod } 4\\ \non
\{n_k^{(1)}\}&=&(3,0,2,2)\hskip 1cm \mbox{diag}(\exp{\frac{2\pi i}{4}}(0,0,0,2,2,3,3)) \hskip 1cm  \sum kn_k^{(1)}\equiv 2 \mbox{ mod } 4\\ \non
\{n_k^{(2)}\}&=&(0,2,2,3)\hskip 1cm \mbox{diag}(\exp{\frac{2\pi i}{4}}(1,1,2,2,3,3,3)) \hskip 1cm  \sum kn_k^{(2)}\equiv 3 \mbox{ mod } 4\\ \non
\{n_k^{(3)}\}&=&(2,2,3,0)\hskip 1cm \mbox{diag}(\exp{\frac{2\pi i}{4}}(0,0,1,1,2,2,2)) \hskip 1cm \sum kn_k^{(3)}\equiv 0 \mbox{ mod } 4 ~,
\eea
where  $k=0, \ldots , q-1$. 
It is no accident that exactly one of these four has determinant equal to unity (the fourth one), making it an element of $SU(7)$. More generally, for a given partition $\{n_k^{(0)}\}$, consider the following set of $q$ distinct partitions:
\[ \{n_k^{(j)}\}=\{n^{(0)}_{k+j}\} \hskip 1cm j,k=0, 1, \ldots , q-1~,\]
(indices are understood mod $q$). The determinant of the corresponding matrix is given by
\[ D_{\{n_k^{(j)}\}}=\exp{\frac{2\pi i}{q}}(\sum _{k=0}^{q-1}kn_k^{(j)})=\exp{\frac{2\pi i}{q}}(\sum _{k=0}^{q-1}kn_{k+j}^{(0)}) ~.\]
 As we cycle through $j=0, \ldots , q-1$, the sums differ from each other by multiples of $\sum_k n_k^{(j)} = p$:

\bea \non D_{\{n_k^{(j)}\}}-D_{\{n_k^{(j+1)}\}}&=&\sum _{k=0}^{q-1} ( kn_k^{(j)}-kn_k^{(j+1)}) 
\\ \non &=&\sum _{k=0}^{q-1} kn_{k+j}^{(0)}-\sum _{k=1}^{q} (k-1)n_{k+j}^{(0)}
= -qn_{q+j}+\sum_{k=1}^qn_{k+j}^{(0)}  \equiv p \mbox{ mod } q~.
\eea
Since $(q,p)=1$, by the Chinese remainder theorem, the sum 0 mod q, which represents a matrix in $SU(p)$ and not just $U(p)$, appears exactly once. Therefore, to get the number of conjugacy classes in $SU(p)$ we must divide our previous formula by $q$:

\be \label{qinsupder}\frac{1}{q}{p+q-1\choose p }=\frac{(p+q-1)!}{p!\, q!}~. \ee 
Eqn. (\ref{qinsupder}) represents the number of semi-classical vacua appearing on the moduli space of M-theory compactifications on a cone over  $(\S ^3 \times \S ^3)/(\Z_p\times \Z_q)$, as well as the number of inequivalent ways to break $SU(p)$ gauge symmetry by a $\Z_q$ Wilson line. 

As explained in the introduction, in addition to the total number of ways to break the gauge symmetry by Wilson lines, which is given by Eqn. (\ref{qinupder}) for $U(p)$ and Eqn. (\ref{qinsupder}) for $SU(p)$, it is also of physical interest to count how many of these symmetry breakings  result in a specified number of $U(1)$ factors in the remaining gauge group $H$. Let us denote this number of $U(1)$ factors by $s$, so that the symmetry breaking is of the form

\[ U(n)\longrightarrow  U(1)^{s}\times\Pi _{i=1}^s SU(n_i)  \]
or
\[ SU(n)\longrightarrow U(1)^{s}\times \Pi _{i=1}^{s+1} SU(n_i) .\]

Again we show how to count through an example. We start with the full unitary group $U(p)$. Let $p=7$ and $q=4$ as before, and let $s=2$. Consider the following diagram:

\[ \bullet  \bullet  \hskip -.1cm | \hskip -.1cm \bullet  \bullet  \bullet  \bullet   \hskip .1cm \bullet \]
It corresponds to a partition of $p$ into $s$ non-zero parts, $\{n_a \}=(n _1, n _2)=(2,5)$ ($\sum_{a=1}^s n _a=p$). Note that there is $1=s-1$ vertical dividing line. There are
\[ {p-1\choose  s-1 } \]
different such partitions, also known as "$s$-compositions of $p$" \cite{stanley}. To associate such a partition to a matrix, we must choose $s=2$ eigenvalues out of the possible $q=4$. There are
\[  {q\choose s } \]
different such choices. Suppose we chose eigenvalues $e^{2\pi i/4}$ and $e^{6\pi i/4}$, which we denote by $\{\lambda _a\} = (\lambda _1, \lambda _2 ) = (1,3)$.  Then the diagonal matrix corresponding to this choice is diag $(\exp(2\pi i/4)(1,1,3,3,3,3,3))$, where as before we have ordered the exponents in nondecreasing order so as not to overcount. Note that we have not required here that $p$ and $q$ be relatively prime. So  we have that the number of ways to break the gauge group $U(p)$ into subgroups that have exactly $s$ $U(1)$ factors is 
\be \label{qsinupder} { p-1\choose s-1 } {q\choose s } = \frac{s}{p} {p\choose  s }{q\choose s}~.\ee

The calculation is modified for $SU(p)$, where we impose the condition that the determinant of the matrix is unity and $(p,q)=1$. Continuing with the example $\{n_a \}=(2,5)$, consider the following set of $q=4$ choices of eigenvalues:
\bea \non
\{\lambda_a ^{(0)}\}&=&(1,3)\hskip 1cm \mbox{diag}(\exp{\frac{2\pi i}{4}}(1,1,3,3,3,3,3)) \hskip 1cm  \sum n _a \lambda _a ^{(0)} \equiv 1 \mbox{ mod } 4\\ \non
\{\lambda_a^{(1)}\}&=&(2,0)\hskip 1cm \mbox{diag}(\exp{\frac{2\pi i}{4}}(2,2,0,0,0,0,0)) \hskip 1cm \sum n _a \lambda _a ^{(1)} \equiv 0 \mbox{ mod } 4\\ \non
\{\lambda_a^{(2)}\}&=&(3,1)\hskip 1cm \mbox{diag}(\exp{\frac{2\pi i}{4}}(3,3,1,1,1,1,1)) \hskip 1cm \sum n _a \lambda _a ^{(2)} \equiv 3 \mbox{ mod } 4\\ \non
\{\lambda_a^{(3)}\}&=&(0,2)\hskip 1cm \mbox{diag}(\exp{\frac{2\pi i}{4}}(0,0,2,2,2,2,2)) \hskip 1cm \sum n _a \lambda _a ^{(3)} \equiv 2 \mbox{ mod } 4 ~,
\eea
where $a=1, \ldots ,s$ and 
\[ \{\lambda_a^{(j)}\}=\{ \lambda _a^{(0)} +j\} \hskip 1cm j=0, 1, \ldots , q-1 ~, \]
and all numbers are understood mod $q$. As before, 
 it is no accident that exactly one of these (the second in this case) has determinant equal unity. 
The determinants  of the corresponding matrices are given by 
\be \label{dets} D_{ \{\lambda _a ^{(j)}\} } = \exp {\frac{2\pi i}{q}}(\sum _{a=1}^{s}n _a\lambda_a^{(j)})= \exp {\frac{2\pi i}{q}}(\sum _{a=1}^{s}n _a(\lambda_a^{(0)}+j)~.\ee
As we cycle through $j=0, \ldots , q-1$, the sums differ from each other by multiples of $ p$:

\[ D_{ \{\lambda _a ^{(j+1)}\} } -D_{ \{\lambda _a ^{(j)}\} } =\sum _{a=1}^{s}n _a\lambda_a^{(j+1)} - \sum _{a=1}^{s}n _a\lambda_a^{(j)} =  \sum _{a=1}^s n_a =p~.\]
When $p$ and $q$ are relatively prime, exactly one of the $q$ sums in equation (\ref{dets}) will have determinant equal unity. So we divide Eqn. (\ref{qsinupder}) by $q$ to obtain

\be \label{qsinsupder} \frac{1}{q} { p-1\choose s-1} { q\choose s}  = \frac{s}{pq} { p\choose s }{ q\choose s }~, \ee
which holds for $(p,q)=1$ and represents
 the number of ways to break the gauge group $SU(p)$ into subgroups  that have exactly $s-1$ $U(1)$ factors. Each such breaking also represents a semiclassical point on the branch ${\cal N} _{s, \Gamma}$ of the moduli space associated with $\Gamma = \Z_p\times \Z_q$. 

Note the symmetry under exchange of  $p$ and $q$  in both Eqn.  (\ref{qinsupder}) and Eqn. (\ref{qsinsupder}). 
This symmetry implies a certain duality between the corresponding physical theories, see \cite{Friedmann:2002ct}. 

Putting together Eqns. (\ref{qinsupder}) and (\ref{qsinsupder}) we now obtain 
\[ { p+q-1\choose p }=\sum _{s=1}^p \frac{s}{p} { p\choose s } { q\choose s } =\sum_s { p-1\choose s-1 } {q\choose q-s }.\]
This is known as the Chu-Vandermonde identity \cite{stanley}. This seems to be the first appearance of this identity in the context of counting string/M theory vacua, and its first appearance in the context of counting conjugacy classes of elements of finite order in a group G. 

It is also possible to obtain formulas in the case $(p,q)\neq 1$ via a generating function approach \cite{mathy}. 
For integers $n$ and $m$ that are not relatively prime, the number of conjugacy classes of elements of order $m$ or dividing $m$ in $SU(n)$ is 
\[ {1\over m}\sum_{d|(n,m)}\phi(d) {n/d+m/d  -1 \choose n/d}, \]
where $\phi(d)$ is Euler's function. This is also the number of ways to break $SU(n)$ gauge theory to a subgroup using a Wilson line of order $m$ or dividing $m$. When we require that the remaining subgroup of $SU(n)$ has exactly $s-1$ $U(1)$ factors, the number is
\[ {1\over m}\sum_{d|(n,m)}\sum _{j\geq 0}\phi(d) {n/d+m/d - j -1 \choose n/d-j}{m/d\choose j}{jd\choose s} (-1)^{j+s}.
\]
It remains to be seen how this case of $(p,q)\neq 1$ would arise in the context of M-theory compactifications.

\section{Counting vacua associated with symplectic groups}\label{symplectic}

Here we derive the expressions analogous to those of Section \ref{unitary}, with unitary groups replaced by symplectic groups. 

Every element in the symplectic group $Sp(n)=Sp(n,\C)\cap U(2n)$
is conjugate to an element in the maximal torus $T_{Sp(n)}$ given by
\[ \label{spntorus} T_{Sp(n)}=\left \{ (e^{ i\theta _1}, \ldots ,  e^{ i\theta _n},  e^{- i\theta _1}, \ldots ,  e^{ i\theta _n}) \right \}. \]
We need to count the number of inequivalent such elements $x$ for which $x^m=1$.  Note that the first $n$ diagonal entries determine the entire matrix. Also note that given a matrix in $T_{Sp(n)}$, replacing any $\theta _j$ by $2\pi -\theta _j$ results in an equivalent (i.e. conjugate) matrix. Therefore, we may restrict our attention to the first $n$ entries, and to $\theta _j = \frac{2\pi j}{m}$, where $j=0,1, \ldots [\frac{m}{2}]$ .

The arguments leading to Eqn. (\ref{qinupder}) apply here with $m$ replaced by $\left ( \left [{m\over 2} \right ]+1 \right )$, which is the number of possible entries on the diagonal, and with $p$ replaced by $n$. We get

\be \label{qinspnder}{ n+\left [\frac{m}{2}\right ]\choose n } ~\ee
for the number of conjugacy classes of elements of order dividing $m$ in $Sp(n)$. 

In analogy with Eqn. (\ref{qsinupder}) we get 
\be \label{qsinspnder} { n-1\choose s-1 }{ [\frac{m}{2}]+1\choose s } ~\ee
for the number of conjugacy classes of elements of order dividing $m$ with $s$ distinct conjugate pairs of eigenvalues, that is, $s$ distinct  $\theta _j$ in Eqn. (\ref{spntorus}).

\section{Counting vacua associated with orthogonal groups}\label{orthogonal}
In this section, we count the number of elements of order  dividing $q$ in all the orthogonal groups: $SO(2n)$, $SO(2n+1)$, $O(2n)$ and $O(2n+1)$. We discuss the odd and even values of $q$ separately. The case of $SO(2n)$ corresponds  to singularities of type $D$, and it counts semiclassical vacua on the moduli space of M-theory compactifications on singular $G_2$  spaces asymptotic to a cone over $(\S ^3 \times \S ^3)/{\bf D}_n\times \Z_q$.

\subsection{Elements of odd order}
Any matrix in $SO(2n+1)$ is conjugate to an element of its maximal torus, given by 
\be \label{so2n+1tor}
T_{SO(2n+1)}=\left \{ \mbox{diag} (A(\theta _1), A(\theta _2), \ldots , A(\theta _n), 1 ) \;\;  \;  | \; \; \; A(\theta)=\left ( \begin{array}{cc} \cos \theta & \sin \theta \\ -\sin \theta & \cos \theta 
\end{array} \right )\right \}~.
\ee
Note that the matrix
\[
 \left ( \begin{array}{c|c|c} B &&  \\  \hline & I_{2n-2}& \\ \hline && -1
\end{array} \right )~,
\]
where
\[
B=\left ( \begin{array}{cc} 1 &  \\  & -1 
\end{array} \right ) ~,
\]
is an element of $SO(2n+1)$ and conjugating by it takes $\theta _1$ to $-\theta _1$ in $T_{SO(2n+1)}$. Similarly, replacing any $\theta _j$ by $-\theta _j$ in the expression for $T_{SO(2n+1)}$ leaves us in the same conjugacy class. Therefore, similar to the symplectic case, we may restrict our attention to $\theta _j = \frac{2\pi j}{q}$, $j =0,1, \ldots \left [ {q\over 2} \right ]$. As in Eqn. (\ref{qinupder}), and now replacing $p$ by $n$ and $q$ by $ {q+1\over 2} $, we have
\be \label{oddqinoddsoder} { n+ {q+1\over 2}  \choose n } \ee
for the number of conjugacy classes of elements of order dividing $q$ in $SO(2n+1)$ for odd $q$. Physically, this represents the number of ways to break an $SO(2n+1)$ gauge symmetry by a Wilson line of order $q$, where $q$ is odd.

The maximal torus of $O(2n+1)$ is different:
\be \label{o2n+1tor}
T_{O(2n+1)}=\left \{ T_1= \mbox{diag} (A(\theta _1), A(\theta _2), \ldots , A(\theta _n),  1 ) \; ; \; T_2= \mbox{diag} (A(\theta _1), A(\theta _2), \ldots , A(\theta _n),  -1 )\right \} ~.
\ee
Since here $q$ is odd,  we cannot have $-1$ as an eigenvalue for any element of order $q$, so elements of $T_2$ are not relevant. Therefore, the counting remains the same as the $SO(2n+1)$ case, and Eqn. (\ref{oddqinoddsoder}) holds.

Now consider $SO(2n)$. Here the maximal torus is
\be \label{so2ntor}
T_{SO(2n)}=\left \{ \mbox{diag} (A(\theta _1), A(\theta _2), \ldots , A(\theta _n) )  \right \}~.
\ee
It is no longer the case that replacing $\theta _i$ by $-\theta _i$ necessarily leaves us in the same conjugacy class because 
\[ \left (\begin{array}{c|c}B& 0 \\ \hline 0 & I_{2n-2} \end{array} \right )\notin SO(2n).
\] 
Instead, we must flip the signs of pairs of $\theta _i$'s to stay in the same conjugacy class. To flip, say, $\theta _1$ and $\theta _2$ the conjugation would be via 
\[ \left (\begin{array}{c|c|c}B& 0 & 0\\ \hline 0& B&0\\ \hline 0&0& I_{2n-4} \end{array} \right )\in SO(2n).
\]
so while $\mbox{diag}(A(\theta _1), A(\theta _2), \ldots , A(\theta _n))$ is conjugate to $\mbox{diag}(A(-\theta _1), A(\theta _2), \ldots , A(\theta _n))$ in $O(2n)$, it is not so in $SO(2n)$ unless at least one of the $\theta _j$ is 0. This makes the counting considerablly more complicated in $SO(2n)$ than in $O(2n)$. We start with the easier case, $O(2n)$. 

In $O(2n)$, the maximal torus is 
\be \label{o2ntor}
T_{O(2n)}=\left \{ T_1=\mbox{diag} (A(\theta _1), A(\theta _2), \ldots , A(\theta _n) ) ;   \;  T_2=\mbox{diag}(A(\theta _1), A(\theta _2), \ldots , A(\theta _{n-1}), B ) \right \}~.
\ee
Since $q$ is odd and the order of any element in $T_2$ is even, we can ignore $T_2$ for this case. Now, since 
\[ \left ( \begin{array}{c|c} B&0 \\ \hline 0& I_{2n-2}\end{array}\right )\in O(2n), 
\] 
flipping any single $\theta_j$ leaves us in the same conjugacy class, as was the case for $SO(2n+1)$. So we can restrict to $\theta _j = \frac{2\pi j}{q}$, $j =0,1, \ldots \left [ {q\over 2} \right ] $, giving 
\be \label{oddqinevenoder}{ n+\left [ {q\over 2} \right ] \choose n } ~\ee
for the number of ways to break $O(2n)$ gauge symmetry by a Wilson line of order $q$, where $q$ is odd. 

In going from $O(2n)$ to $SO(2n)$, in some cases two elements that are in the same conjugacy class in $O(2n)$ may be in two distinct conjugacy classes in $SO(2n)$. These are the elements of $T_1$ in the maximal tori of $SO(2n)$ or $O(2n)$ for which $\theta _j\neq 0 \; \forall j$. The number of such elements $x$ with $x^q=1$ is 
\be\label{double}{ n+ {q+1\over 2}  -1\choose n }
~.\ee
This formula is obtained by replacing ${(q+1)\over 2}$ by ${(q+1)\over 2}-1$ in Eqn. (\ref{oddqinoddsoder}), as the number of allowed eigenvalues is one less (since zero is not allowed). 

The number of elements in $O(2n)$ with $x^q=1$ and at least one $\theta _i=0$ is 
\be \label{single} { n-1+{q+1\over 2} \choose n-1}
~.\ee
This formula is derived by fixing $\theta _1=0$ and counting the number of elements in $T_1$ of order dividing $q$, allowing the $\theta _{j>1}$ to take on any of the ${q+1\over 2}$ possible values. It is equivalent to replacing $n$ by $n-1$ in Eqn. (\ref{oddqinoddsoder}). 

To get the number of conjugacy classes in $SO(2n)$ of order $q=2m+1$, we multiply formula (\ref{double}) by 2 and add formula (\ref{single}) to get

\be  \label{oddqinevensoder}
2 { n+ {q-1\over 2} \choose n } + { n+ {q-1\over 2}\choose n-1  } = { n+ {q+1\over 2} \choose n-1 } \frac{n+q-1}{n} ~,
\ee
which physically is the number of ways to break $SO(2n)$ gauge symmetry by Wilson lines of odd order $q$. This case of $SO(2n)$ includes the important GUT group $SO(10)$, which is obtained when the $G_2$ manifold has a $D_5$ singularity, as constructed in \cite{Friedmann:2002ct}.
Note that the formulas for $O(2n)$, $O(2n+1)$, and $SO(2n+1)$ are identical, but differ from the formula for $SO(2n)$. 

Now let $s$ be the number of distinct $\theta _i$ that appear in Eqns. (\ref{so2n+1tor}), (\ref{o2n+1tor}), (\ref{so2ntor}), or (\ref{o2ntor}).

In analogy with Eqn. (\ref{qsinupder}), we get
\[ { n-1 \choose s -1 }{  {q+1\over 2} +1\choose s }
\]
for the number of conjugacy classes of elements of order dividing $q$ in $O(2n)$, $O(2n+1)$, or $SO(2n+1)$ with $s$ distinct $\theta _i$'s. 

For $SO(2n)$, the counting is again more complicated. We count separately the conjugacy classes that have at least one $\theta _i=0$, and those that do not. If the angle $0$ does not appear, the number of elements of order dividing $q$ with eigenvalue number $s$ is
\be  \label{so2nnot0} { n-1 \choose s -1 }{ \left [ {q\over 2}\right ] \choose s }
\ee
and as before we will multiply this by 2. If the angle $0$ does appear,  we have $s-1$ distinct angles left to be determined:
\[  {n-1 \choose s -1 }{ \left [ {q\over 2}\right ] \choose s-1 }~.
\]
So the total number of conjugacy classes of elements of order dividing $q$ in $SO(2n)$ with $s$ distinct $\theta _j$ is
\be  \label{soddqinevensoder} { n-1 \choose s -1 } \left [2 {  {q+1\over 2} \choose s }{  {q+1\over 2} \choose s-1 } \right ]~.
\ee

As in the case of unitary groups, the  number $s$ here has physical significance as it is related to the number of $U(1)$ factors that appear in the remaining group $H$. Given the relation 
\[ U(n)=O(2n)\cap Sp_{2n}\R ~,\]
one can rewrite an element in the even part of $T_{O(2n)}$   as $\mbox{diag}(e^{i\theta _1}I_{n_1}, \ldots , e^{i\theta _s}I_{n_s})\in U(n)$ where $\sum _{j=1}^sn_j = n$ and $n_j$ is the number of times $\theta _j$ appears. Then the commutant of this element in $U(n)$ is 
\[ \prod _{j=1}^s U(n_j)=U(1)^s\;\times  \prod _{j=1}^s SU(n_j) ~,\] 
so in $O(2n)$ the commutant of that element has at least $s$ $U(1)$ factors.

\subsection{Elements of even order}

For $SO(2n+1)$, all the arguments that applied for odd $q$ apply for even $q$ as well. For $O(2n+1)$, since $q$ is even, we do have to add the elements of type $T_2$ in the torus. But the counting is exactly the same as that in $T_1$. So we have 

\be \label{evenqinoddsoder}{ n+\frac{q}{2}\choose n }\ee
for the number of  conjugacy classes of elements of order dividing $q=2m$ in $SO(2n+1)$, and
\be \label{evenqinoddoder}2{ n+\frac{q}{2}\choose n}\ee
for that number in $O(2n+1)$.

Now we turn to $O(2n)$. Again, in addition to the counting in $T_1$, we also have to count inside $T_2$. For $T_2$, note that we have $n-1$, not $n$, $2\times 2$ blocks of the form $A(\theta _i)$. We obtain 
\be \label{evenqinevenoder} { n+\frac{q}{2}\choose n } +{ n+\frac{q}{2} -1\choose n-1 }  ~
\ee
for  the number of conjugacy classes of elements of order dividing $q=2m$ in $O(2n)$.

For $SO(2n)$, the maximal torus consists only of elements of the form $T_1$. However, flipping $\theta _j$ to $-\theta _j$ does not leave us in the same conjugacy class unless at least one $\theta_j$ is either $0$ or $\pi$ so that $A(\theta _j)$ commutes with $B$ (note that in the previous subsection where $q$ was odd, $\theta _j = \pi$ was not allowed). So now we need to count separately the elements in the torus that have no $\theta _j = 0, \pi$, multiply this number by 2, and add to that the number of elements that do have at least one $\theta _j=0,\pi$. Our counting methods, together with some algebra of binomial coefficients, leads to
\be \label{evenqinevensoder}
{ n+\frac{q}{2}\choose n }  +{ n+\frac{q}{2}-2\choose n } ~.
\ee

We now consider the number of conjugacy classes of order $q=2m$ with a fixed  number $s$ of distinct values of $\theta _i$ in the torus. We have
\[ { n-1 \choose s -1}{ {q\over 2}+1 \choose s } \]
for $SO(2n+1)$, 
\[ 2 { n-1 \choose s -1 }{  {q\over 2}+1 \choose s }~\]
for $O(2n+1)$, 
\[ \left [ { n-1 \choose s -1 } +  { n-2 \choose s -1 }\right ]{ {q\over 2}+1 \choose s } ~\]
for $O(2n)$, and 
\[ 
{ n-1 \choose s -1 }\left [ { {q\over 2}+1\choose s } +{  {q\over 2}-1\choose s } \right ]~\]
for $SO(2n)$.

\section{On symmetry breaking patterns: Wilson  vs. Higgs}\label{wilsonvshiggs}

Patterns of symmetry breaking by Wilson lines 
are generally considered to be the same as patterns of symmetry breaking by a related adjoint Higgs field. However, there is a subtle but significant difference which we point out here.

As explained in Section \ref{vacconj}, when a gauge symmetry $G$ is broken by a Wilson line $U$, the remaining group $H$ is given by the centralizer of the Wilson line in the group, 
\[ H=C_G(U)=\{ h\in G \; | \; hUh^{-1}=U \}.\]
On the other hand, in symmetry breaking by a Higgs field, 
 the remaining group is the one generated by the commutant of the Higgs field in the Lie algebra, 
\[ H=\exp \left ( \{ X\in \mathfrak{g} \; | \; [\phi, X]=0 \} \right ). \]

We now show that even with the simple identification $U=e^\phi$ between a Wilson line and a Higgs field, the symmetry breaking patterns are not the same. The difference is rooted in the fact that in general, conjugation in the Lie group is not in one-to-one correspondence with commutation  in the Lie algebra. Specifically, let $X\in \mathfrak{g}$  and let $U_t=e^{t\phi}, h_t=e^{tX}$ form one-parameter subgroups of the Lie group. Then 
\be \label{leftoverwilson} h_tU_th_t^{-1}=U_t \ee
does not always imply
\be \label{leftoverhiggs} [\phi, X]=0 ~.\ee
Rather, this is true only when $t$ is small; when $t$ is not small, (\ref{leftoverwilson}) may be satisfied while (\ref{leftoverhiggs}) is not. We demonstrate this by a simple explicit example. Let  $\phi = \mbox{diag}(\pi i , -\pi i)\in \mathfrak{su}(2)$. Then $U=e^\phi=-I$, which is in the center of the group $SU(2)$ so does not break it. But $\phi$ is not in the center of the Lie algebra $\mathfrak{su}(2)$; it commutes only with its diagonal elements. 
 Hence, the symmetry breaking pattern arising from an adjoint Higgs field can be different from that arising from the corresponding Wilson line - at the group level, more symmetry may be preserved. Thereofore, one should take extra care in applying the formulas obtained here for Wilson lines to symmetry breaking by Higgs fields. 

This allows us to use Wilson lines to study certain GUT symmetry breaking patterns that are not allowed by Higgs fields  and obtain the standard model group in new ways \cite{wip}.

\newpage
\no {\bf Acknowledgments}

The authors are grateful to Jonathan Pakianathan, Steve Gonek, Ben Green, Dan Freed, and Dragomir Djokovic  for discussions. 
The authors are also grateful to Ori Ganor for comments on a draft. The work
of the first author was supported in part by US DOE Grant number
DE-FG02-91ER40685, and of the second author in part by NSF Grant
number DMS-1068625.

\renewcommand{\baselinestretch}{1.2}

\end{document}